\documentclass[aps,pra,showpacs,preprintnumbers,superscriptaddress,twocolumn,nofootinbib]{revtex4-2}
\usepackage{amsmath,amssymb}
\usepackage{bm}
\usepackage{tipa}
\usepackage{upgreek}
\usepackage{comment}
\usepackage{mathrsfs}
\usepackage{graphicx}
\usepackage{braket}
\usepackage{enumitem}
\usepackage{mathtools}
\usepackage{mathbbol}
\usepackage{gensymb}
\usepackage[normalem]{ulem}
\usepackage{color}
\usepackage[colorlinks,bookmarks=true,citecolor=blue,linkcolor=red,urlcolor=blue]{hyperref}
\usepackage{hyperref}
\usepackage{appendix}
\usepackage{xr-hyper}

\renewcommand{\vec}[1]{\mathbf{#1}}
\usepackage{pifont}
\usepackage{todonotes}
\usepackage[inkscapelatex=false]{svg}
\usepackage{subcaption}
\usepackage{ragged2e} % to align the caption to column-width \justifying

\newcommand{\be}{\begin{equation}}
\newcommand{\ee}{\end{equation}}
\newcommand{\bea}{\begin{eqnarray}}
\newcommand{\eea}{\end{eqnarray}}
\newcommand{\nn}{\nonumber}

\DeclareMathSymbol{\shortminus}{\mathbin}{AMSa}{"39}

\usepackage{soul}

\begin{document}

\title{Direct energy dissipation measurements for a driven superfluid via the harmonic-potential theorem}

    \author{Clara~Tanghe}
    \affiliation{Department~of~Physics~and~Astronomy,~Ghent~University,~Krijgslaan~281,~9000~Ghent,~Belgium} 
    \author{Senne~Van~Wellen}
    \affiliation{Department~of~Electronics~and~Information~Systems,~Ghent~University,~Technologiepark-Zwijnaarde~126,~9052 Ghent,~Belgium}
    
    \author{Kobe~Vergaerde}
    \affiliation{Department~of~Electronics~and~Information~Systems,~Ghent~University,~Technologiepark-Zwijnaarde~126,~9052 Ghent,~Belgium}
    \author{Karel~Van~Acoleyen}
    \affiliation{Department~of~Physics~and~Astronomy,~Ghent~University,~Krijgslaan~281,~9000~Ghent,~Belgium}
    \affiliation{Department~of~Electronics~and~Information~Systems,~Ghent~University,~Technologiepark-Zwijnaarde~126,~9052 Ghent,~Belgium}

	\begin{abstract}
    {We propose and experimentally demonstrate a method to directly measure energy dissipation for a linearly driven superfluid confined in a harmonic trap. The method relies on a perturbed version of the harmonic-potential theorem, according to which a potential perturbation - effectively acting as a stirrer - converts center-of-mass motional energy into internal energy. Energy conservation then enables a direct, quantitative determination of the dissipated energy from measurements of the macroscopic center-of-mass observables. Applying this method to a perturbed, driven Bose-Einstein condensate, we observe dissipation curves characteristic of superfluid flow, including a critical velocity that depends on the stirrer strength, consistent with previous studies. Our results are supported by mean-field simulations, which corroborate both the theoretical framework and the experimental findings.}
    \end{abstract}
	
 	\maketitle

\section{Introduction}

Superfluidity is one of the most striking collective phenomena in many-body physics. Although it was first observed in liquid helium, the advent of Bose-Einstein condensates (BECs) in dilute atomic gases has opened an entirely new window into the microscopic nature of this phenomenon. In particular laser stirring experiments represent the most direct probes for superfluidity in such cold atom systems and have been used to study superfluid properties for a wide variety of systems - both bosonic and fermionic -
for different inter-atom interactions, trap dimensions and stirrers \cite{Raman1999, Raman2000, Desbuquois2012,
Wenzel2018,
Onofrio2000,
Miller2007,
Weimer2015,
Sobirey2021,
Madison2000,
Inouye2001,
Neely2010,
Kwon2015,
Kwon2016,
Engels2007,
Lye2005,
Chen2008,
Dries2010,
Ferrier-Barbut2014}.
In those experiments superfluidity is diagnosed in terms of the response to the motion of the optical potential perturbation $\delta V$ (the stirrer) on the purported superfluid state. The superfluid character then manifests itself in the existence of a critical velocity $v_c$, such that there is only considerable energy dissipation for stirring velocities $v$ above this velocity. Often this energy dissipation is observed indirectly, e.g. from the induced changes in the condensate fraction or density profile \cite{Onofrio2000,
Miller2007,
Weimer2015,
Sobirey2021,Engels2007}, the excitation of vortices \cite{Madison2000,
Inouye2001,
Neely2010,
Kwon2015,
Kwon2016} or solitons \cite{Engels2007}. 

Other experiments employ a calorimetric approach by measuring the energy dissipation in terms of the temperature increase \cite{Raman1999, Raman2000, Desbuquois2012,
Wenzel2018}. For this, the atom cloud is held in the trap for some time after the stirring stage, after which a temperature is extracted from time-of-flight (TOF) or in-situ (IS) images. As such, this procedure still does not constitute a true direct measurement of the energy dissipation. But more importantly, it relies crucially on the assumption that during the hold stage the system equilibrates to a (local) thermal state, while thermalization, especially for dilute weakly-interacting systems, can be a slow process. In fact, for the case of the historic first BEC superfluidity experiment \cite{Raman1999}, recent numerical and analytical work  indicates that this thermalization assumption was indeed wrong \cite{Kiehn2022}. 

Here we propose and experimentally demonstrate a new method for directly measuring the energy dissipation in a stirring experiment, relying on energy conservation rather than thermalization. The method hinges on the harmonic-potential theorem (HPT) \cite{Dobson1994}. As we show below, a perturbed version of this theorem implies that the total energy of a linearly driven quantum many-body system in a harmonic trap supplemented by a static perturbation $\delta V$, naturally decomposes into a part $e_{COM}$ associated to the center-of-mass (COM) motion and a part $e_{int}$ corresponding to the energy in the COM rest frame. Energy conservation then allows the determination of the latter internal energy evolution from the damping of the harmonic COM motion, as measured from the readily accessible COM position and velocity of the atom cloud.  The perturbation $\delta V$ acts as a moving stirrer in the COM rest frame and it is precisely this relative stirring which excites the internal degrees of freedom. Our procedure therefore indeed yields a direct measurement of the energy dissipation due to stirring. It is quite literally the quantum analogue of the energy dissipation and friction force measurements of a classical moving object (e.g. a bike) from the evolution of its COM motion (e.g. the bike's deceleration when the cyclist stops pedaling).

Note that previous stirring experiments already probed superfluidity by studying the damping of the harmonic motion due to a potential perturbation \cite{Lye2005,
Chen2008,
Dries2010,
Ferrier-Barbut2014}. The novelty of our approach lies in the use of the HPT to turn the observed damping into a quantitative measurement of the instantaneous energy dissipation. Furthermore, in these earlier experiments the COM degrees of freedom were excited by a sudden shift of the trap minimum, whereas our linear driving protocol ramps up the energy in a gradual manner. We should also mention the previous theoretical work \cite{Wu2011,Wu2014} on a perturbed version of the HPT. The main difference with our approach is that we work in the actual COM rest frame rather than the unperturbed COM rest frame.

\section{Harmonic-potential theorem} To write down the perturbed version of the HPT, we start from the (lab frame) Hamiltonian in second quantization, that describes a general uniformly interacting (bosonic or fermionic) non-relativistic many-body system in a perturbed harmonic trap, subject to a linear drive:
  \be  H=  H_0 +  H_{\delta}+ H_D(t) \,,\label{eq: H}\ee
  where: \bea  H_0 &=& \int\hspace{-0.13cm} d^3x\, \frac{\hbar^2}{2m}{ \bf \nabla}\psi^\dagger({\bf x}).{\bf \nabla} \psi({\bf x}) +\frac{m\omega_i^2x_i^2}{2}\psi^\dagger({\bf x})\psi({\bf x})\nn\\&&+\int\hspace{-0.13cm} d^3xd^3y \frac{1}{2}V_I({\bf x}-{\bf y})\psi^\dagger({\bf x})\psi^\dagger({\bf y})\psi({\bf x})\psi({\bf y})\,,\\
  H_{\delta} &=&\int\hspace{-0.13cm} d^3x\,\delta V({\bf x})\psi^\dagger({\bf x})\psi({\bf x})\,,\\ H_{D}&=&-\int\hspace{-0.13cm} d^3x\,{\bf f}(t).{\bf x}\,\psi^\dagger({\bf x})\psi({\bf x})\,.\label{eq:HD}\eea

  The COM position and momentum observables are defined as:
\be \hat {\bf r}=\frac{1}{N}\int\hspace{-0.13cm} d^3x\,{\bf x}\,\psi^\dagger\psi\,,\quad\quad  \hat {\bf p}=\frac{1}{N}\int\hspace{-0.13cm}d^3x\,\psi^\dagger(-i\hbar {\bf \nabla})\psi\,, \label{eq:defrp}\ee
with $N=\int\! d^3x\,\psi^\dagger\psi$ the particle number. Notice that both these observables are intensive, in particular we have for the total momentum $\hat {{\bf P}}=N \hat {\bf p}$. Due to their intensive nature these observables behave classically for large particle numbers: e.g. $\braket{\hat r_i \hat r_j}=\braket{\hat r_i}\braket{\hat r_j}+\mathcal{O}(1/N)$. 

For the time-evolution of the Hamiltonian and COM observables we then have (see also \cite{Wu2011,Wu2014}):
\bea \dot {\hat r}_i&=&\frac{i}{\hbar}[H,\hat r_i]=\frac{\hat p_i}{m}\,,\label{eomr}\\
\dot {\hat p}_i&=&\frac{i}{\hbar}[H, \hat p_i]=-m\omega_i^2  \hat r_i+f_i-\int\hspace{-0.13cm} \frac{d^3x}{N}\,\partial_i(\delta V)\psi^\dagger\psi\,, \label{eomp}\\
\dot H&=&- N \dot f_i  \hat r_i\,.\label{eomH}\eea
In absence of the perturbation, $\delta V=0$, the equations for the COM position and momentum close and correspond to the undamped driven harmonic oscillator. This shows the protection of the harmonic dipole mode under the inclusion of the interactions, sometimes referred to as the generalized Kohn theorem, first demonstrated in \cite{Halperin89}. The extra term that appears in the momentum-evolution when the perturbation is nonzero, couples the COM motion to the internal degrees of freedom and is readily interpreted as a frictional force per particle. For $\delta V\neq 0$ the drive ${\bf f}(t)$ will therefore not only excite the COM degrees of freedom, part of the driving energy will also excite the system internally. 

To quantify this in terms of energy, we first write down the Hamiltonian $H'$ in the COM rest frame. Going from the lab frame to the COM rest frame amounts to applying a coordinate transformation ${\bf x}\rightarrow {\bf x'}={\bf x}-{\bf r}(t)$, with ${\bf r}(t)=\braket{\hat {\bf r}}_t$, the COM position for the particular time-dependent state at hand. For the corresponding field transformation we write:
\be \psi'({\bf x'},t)=e^{-\frac{i}{\hbar} \left(m {\bf v}.{\bf x}' +S(t)\right)}\psi({\bf x},t)\,,\label{phasefac}\ee with ${\bf v}$ the COM velocity, ${\bf v}(t)=\dot {\bf r}(t)=\braket{\hat {\bf p}}_t/m$ and $S(t)$ the classical action in the unperturbed trap: \be S(t)=\int^t\hspace{-0.13cm }du\,\left(\frac{1}{2} m\vec{v}^2-\frac{1}{2}m\omega_i^2r_i^2+{\bf f}.{\bf r} \right)\,. \ee  The COM Hamiltonian $H'$ - i.e. the operator that generates the time-evolution of $\psi'$ - written in terms of the COM rest frame coordinates ${\bf x}'$ and field $\psi'$ then reads (see \textcolor{red}{}):
\be  H'=  H'_0 +  H'_{\delta}(t)+  H'_D(t) \,,\label{Hcom}\ee
with $ H'_0$ of the exact same form as $ H_0$, $ H'_{\delta}$ now displaying a time-dependent potential perturbation that tracks the COM motion: \be \delta V'({\bf x}',t)=\delta V({\bf x}'+{\bf r}(t))\,, \ee and the linear drive in $H'_D$ reading: \be {\bf f}'(t)=\frac{1}{N}\int \hspace{-0.13cm} d^3 x\, \braket{({\bf \nabla}\delta V)\psi^\dagger\psi}_t=\frac{1}{N}\int \hspace{-0.13cm} d^3 x' \,\braket{({\bf \nabla}'\delta V'){\psi'}^\dagger\psi'}_t \,.\ee 

The first two terms in the COM Hamiltonian describe a static harmonic trap with a moving stirrer $\delta V'$. The drive $H'_D$ provides an extra uniform force that cancels out the deceleration due to the frictional force (third term in Eq. \eqref{eomp}), but note that due to the COM rest frame condition, ${\bf 0}={\bf r}'=\braket{\int d^3 x' {\bf x}'{\psi '}^\dagger \psi'}_t/N$ we have $\braket{H'_D}_t=0$.

Finally we can write down the energy conservation relation that separates the total energy per particle in the lab frame $e=\braket{H}_t/N$ into a part associated to the COM motion and a part that corresponds to the energy in the COM rest frame, $e_{int}=\braket{H'}_t/N=\braket{H_0'+H'_\delta}_t/N$. In Appendix \ref{sec:A1}, we show that: \be e=\left(\frac{1}{2}m {|\bf v|}^2+\frac{1}{2}m \omega^2_i r^2_i-{\bf f}.{\bf r}\right) +e_{int}=e_{COM}+e_{int}\,,\label{econs}\ee with $\dot e=-\dot{\bf f}.{\bf r}$, from Eq. \eqref{eomH}. 
These are the two key relations that we use in our experiment to infer the dissipation of the energy related to the COM degrees of freedom into the internal excitation energy. Remark that the corresponding frictional force, proposed as a diagnostic for (non-)superfluidity in \cite{Onofrio2000,Pavloff2002}, follows from $\dot{e}_{int}={\bf f}'(t).{\bf v}$.   

\begin{figure}
    \centering
    \includegraphics[width=0.9\columnwidth]{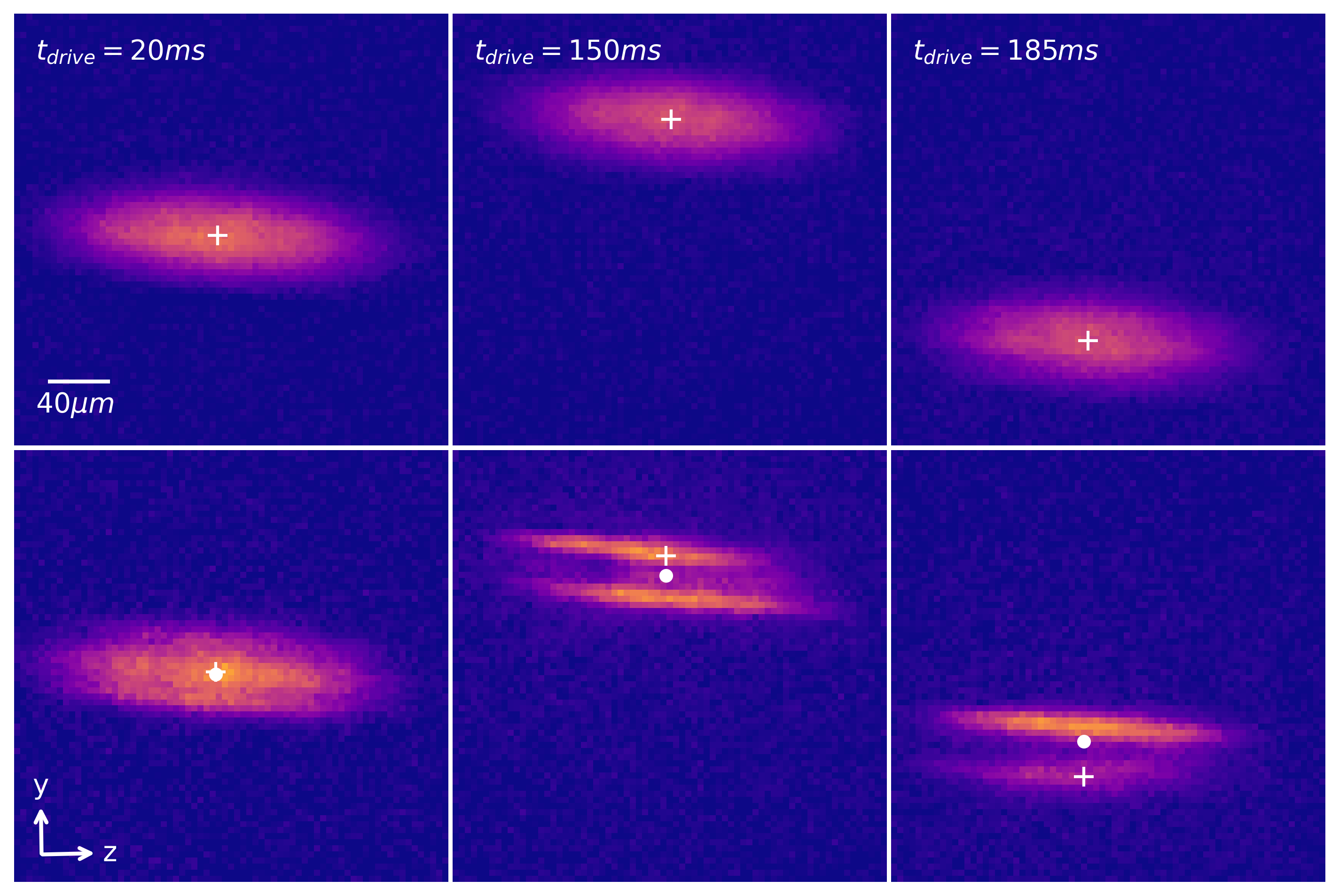}
    \caption{
    \justifying
    {\bf Time-of-flight resonant absorption images} of both the unperturbed (first row) and perturbed (second  row) BEC for different drive times ($t_{hold}=0$, see main text). The center of mass position of the unperturbed (perturbed) cloud is indicated by a cross (dot).}
    \label{fig: tofimages}
\end{figure}

Note also that in the zero-perturbation case $\delta V=0$, the frictionless harmonic COM motion implies that there should be no dissipation into internal energy. For the undamped driven harmonic motion we have indeed $\dot e=\dot e_{COM}$, which is consistent with the time-independence of the COM Hamiltonian $\dot H'=\dot H'_0=0$. This is the original HPT \cite{Dobson1994}, which implies that in the COM rest frame, the groundstate (or any other eigenstate) will not evolve. Experimentally, this manifests itself as an atom cloud undergoing undamped (driven) oscillations, without changing its internal shape if it started off in the groundstate at $t=0$, see e.g. the first row of \autoref{fig: tofimages}. \vspace{0.1cm}

\section{Experimental results} Our experiments start from a nearly pure BEC ($N_0/N\gtrsim 90\%$) of $N_0\approx55\,000$ $^{87}$Rb atoms ($\ket{F = 2, m_F=2}$) in an elongated cigar-shaped magnetic trap, with trapping frequencies $(\omega_x,  \omega_y, \omega_z) = 2\pi \times (440(2), 41.6(2), 440(2))$~Hz (see Appendix \ref{section: experimental set-up} for details on our experimental set-up). On top of the harmonic trap, we apply an optical potential perturbation $\delta V$, which is provided by a blue detuned laser beam shaped by two perpendicular acousto-optical deflectors. Specifically, to a good approximation the potential due to our projected beam reads:  \be \delta V(y,z)= \delta V_0 \frac{w_0}{w(z)}e^{-\frac{2 y^2}{w(z)^2}}\,, \label{eq:pot}\ee with $w(z)=w_0\sqrt{1+(z/z_R)^2}$, where $w_0 = 1.2$~$\mu$m and $z_R = 3.7$~$\mu$m. We will show experimental results for five different nonzero beam intensities, that we will quote in terms of the resulting central potential strengths $\delta V_0$  relative to the chemical potential of the unperturbed BEC, $\mu_0=150$~nK. Note that similar to the Gaussian perturbation used in the harmonic trap dissipation experiment of \cite{Dries2010}, our $\delta V$ crosses the entire harmonic trap orthogonal to the long ($y$-) axis at its center (for our cloud the Thomas-Fermi radii read: $R_{x,z}\approx 2.0$~$\mu$m, $R_y\approx 21$~$\mu$m). Therefore we expect the stirrer to excite solitons and (longitudinal) phonons rather than vortices, which is confirmed by our mean-field simulations, see \autoref{sec:mean-field}. 

Furthermore, we subject the atom cloud to a time-dependent linear magnetic gradient field in the longitudinal $y$-direction, generated by an external coil with a function generator supplying a sinusoidal current with the same frequency as the longitudinal frequency $\omega_y$. 
This resonantly excites the COM motion in the $y$-direction.
Since this drive is far off resonance with respect to the radial trapping frequencies, the effect on the COM motion in the radial directions is negligible, even in case of some slight misalignment of the gradient field. 
We indeed observe no COM oscillations in these directions and for all purposes we can write for the resulting drive term in the Hamiltonian (see \eqref{eq:HD}): \be {\bf f}(t)=(0,f(t),0) = (0,\kappa \sin (\omega_y t),0)\,, \label{driveex}\ee with $\kappa$ proportional to the current amplitude.

Our experimental sequence then goes as follows (see also Appendix \ref{sec: Position and velocity measurements}). Starting from the BEC groundstate in the harmonic trap, we first linearly ramp on the optical potential $\delta V$ during $t_{ramp}=20$~ms. After this, we turn on the drive for a variable time $t_{drive}$ between 5~ms and 200~ms. Then, we turn off both the drive and the optical potential, to let the cloud evolve in the harmonic trap for six different hold times, $t_{hold}=0$~ms - 25~ms in 5~ms increments.
 Finally, we also turn off the harmonic trap, to image the atom cloud in the $YZ$-plane after 24~ms time-of-flight.

\autoref{fig: tofimages} shows examples of resulting TOF images of the (shape-inverted) cloud, at three different driving times (with $t_{hold}=0$~ms), for the unperturbed case  - with $\delta V_0=0$ throughout the whole sequence - and a perturbed case. One can clearly see the unperturbed HPT in action, with the unperturbed cloud shaking in the $y$-direction, without altering its shape. In the perturbed case, the cloud first tracks the unperturbed motion, while at later times it lags behind and its shape is altered considerably. This is a direct manifestation of the conversion of COM energy into internal energy, according to the HPT energy conservation relation \eqref{econs}.

\begin{figure}[t]
    \centerline{\includegraphics[width=1.05\columnwidth]{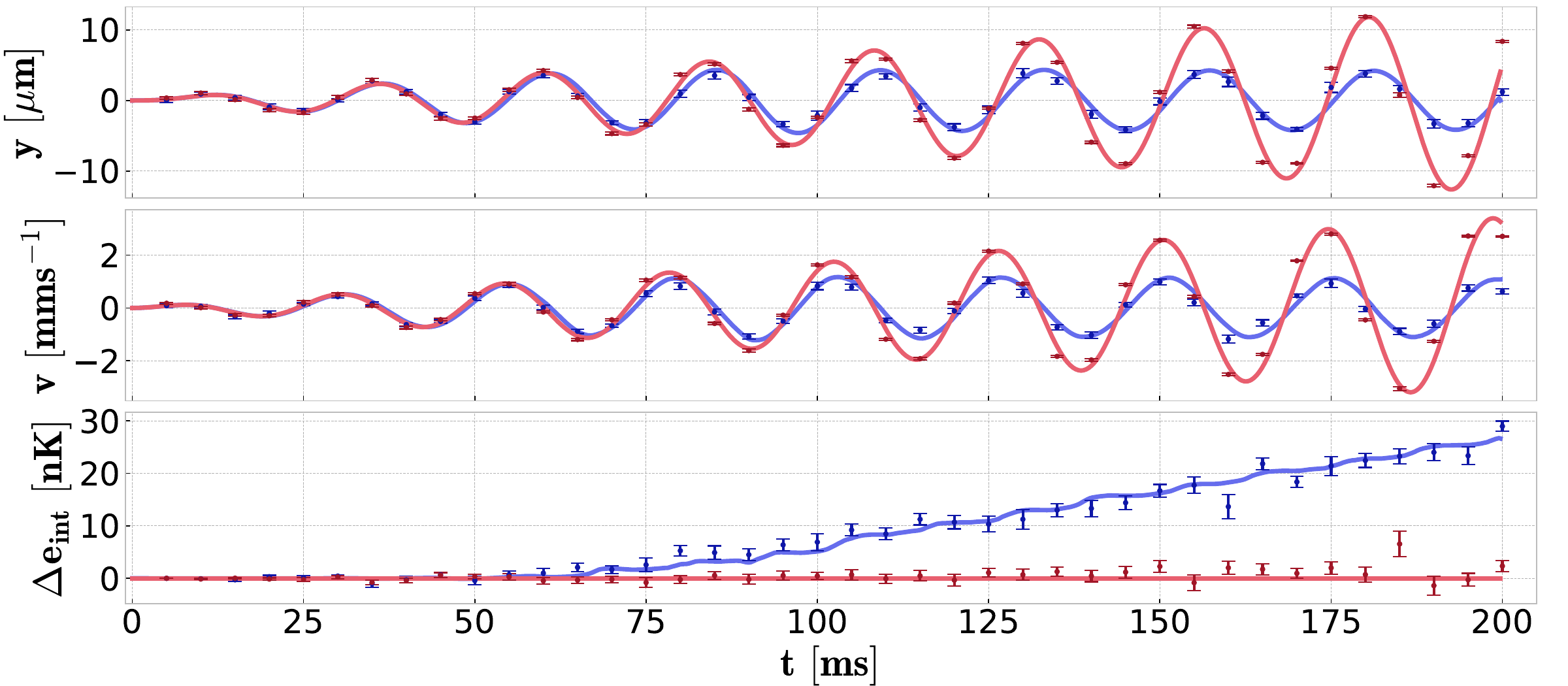}}
\caption{\justifying\textbf{Center of mass motion and internal energy evolution}. 
First two panels: the COM positions $y$ and COM velocities $v$ for the drive of Eq. (\ref{driveex}). Third panel: the inferred internal energy evolution, via Eq. (\ref{intenergyexp}). Red data-points show our experimental results for the unperturbed case ($\delta V=0$), the red lines show the corresponding theoretical prediction of undamped harmonic motion, Eq. (\ref{eomunpert}). Blue data-points show our experimental results for a perturbed case, $\delta V=0.27\mu_0$. The blue lines show the results from our corresponding mean-field simulations (see \autoref{sec:mean-field}).}
\label{fig:data1}
\end{figure}

The quantitative determination of the dissipation $\Delta e_{int}(t)=e_{int}(t)-e_{int}(0)$ from Eq.\eqref{econs} requires accurate measurements of the in-situ COM position $y$ and velocity $v$ in the $y$-direction as a function of the drive time. To enhance the resolution on the in-situ motion we use time-of-flight images. In Appendix \ref{sec: Position and velocity measurements}, we show how to extract both in-situ position and velocity, from TOF measurements at different values of $t_{hold}$, employing the unperturbed HPT. \autoref{fig:data1} displays our results for the unperturbed case and for a perturbed case. Note that every $(y,v)$ data-point in this plot is the result of 18 (or 36) destructive TOF measurements. As shown in the upper panels the unperturbed results for the COM motion fit very well to the unperturbed HPT prediction of an undamped driven harmonic motion (from Eqs.\eqref{eomr} and \eqref{eomp}):
\be
    y(t) = \frac{\kappa}{2m\omega^2}\left(\sin(\omega t)-t \omega\cos(\omega t)   \right)\,, \quad v(t)= \frac{\kappa}{2m} t\sin(\omega t) \label{eomunpert}
\,,
\ee
with $\omega=\omega_y$. The single-parameter fit allows us to determine the value of $\kappa= 0.359(1)$~nK/$\mu$m. We have used the very same drive amplitude for all the perturbed traps. For the perturbed data the COM motion, which initially coincides with the unperturbed motion, shows damping at later times. 

Finally, the lower panel shows the internal energy evolution, as extracted from the COM position and velocity data $(y(t_i),v(t_i))$, by using the HPT energy conservation relation (\ref{econs}) for a driven COM motion in the $y$-direction:
\bea \Delta e_{int}(t)&=& \Delta e(t)-\Delta e_{COM}(t) \label{intenergyexp}\\&=&-\int^t_0 \hspace{-0.13cm} du\,(\dot f y)(u) -\left(\frac{1}{2}m v^2-\frac{1}{2}m \omega^2 y^2+fy\right)\,. \nn\eea
Here, to evaluate from our data the integral expression for the total energy $\Delta e(t)$, we use the Simpson's 1/3 rule approximation, see Appendix~\ref{sec: Simpson} for an estimate on the induced error. The results for the unperturbed case are consistent with the unperturbed HPT result $\Delta e_{int}=0$. In the perturbed case we find, as anticipated from the COM motion, that at early times the internal energy stays approximately zero up to a certain time $t_{onset}\approx 50$~ms after which the internal energy starts to grow. 

\begin{figure}[t]
\centerline{\includegraphics[width=1.0\columnwidth]{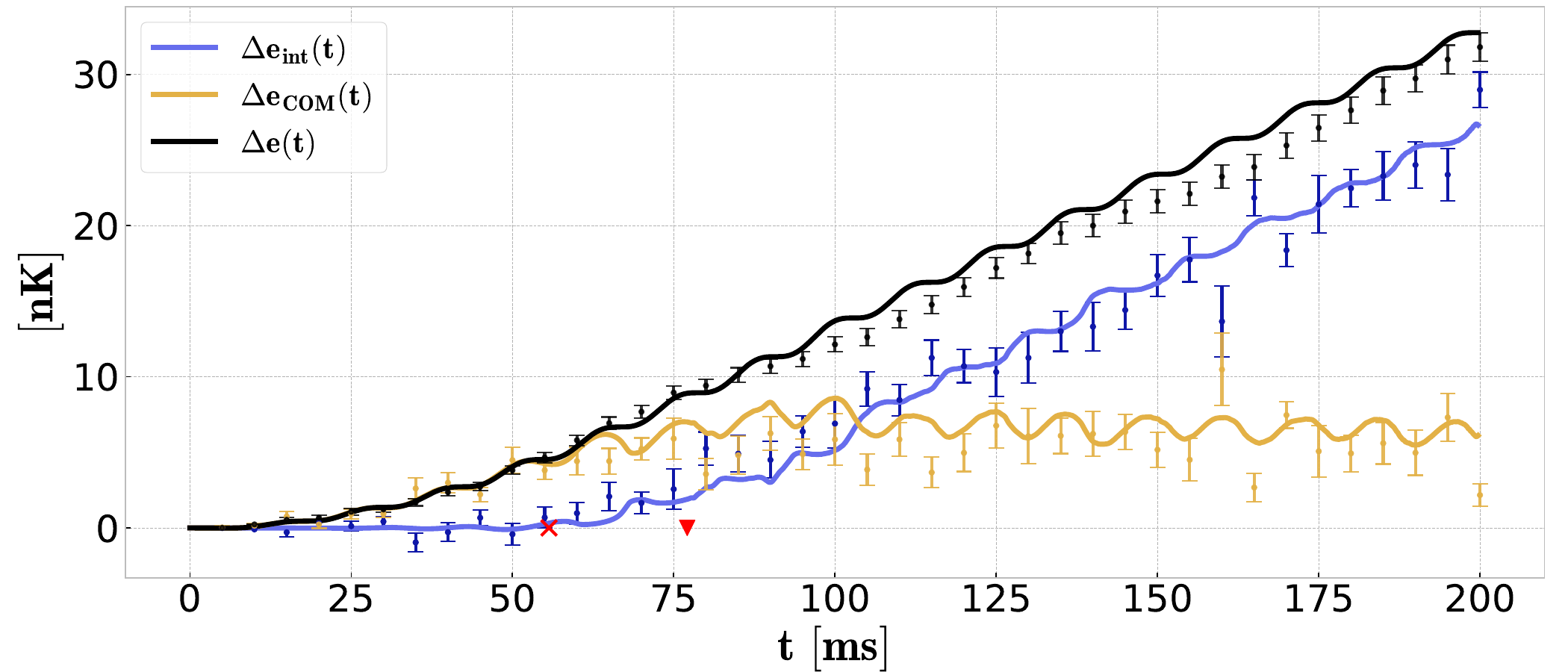}}
    \caption{
    \justifying
    \textbf{Energy curves.} For $\delta V_0=0.27 \mu_0$, the evolution of the total energy $\Delta e$ (black), the COM energy $\Delta e_{COM}$ (yellow) and the internal energy $\Delta e_{int}=\Delta e-\Delta e_{COM}$ (blue), inferred from the COM positions and velocities (see Eq. (\ref{intenergyexp})). The data-points show our experimental results, the lines show the results from our mean-field simulations (see \autoref{sec:mean-field}). Red triangle (cross): estimate of the onset time $t_{onset}$ of dissipation based on the local Landau criterion, from Eq. (\ref{vcrit}) (from our mean-field simulations). 
    }
    \label{fig:Energies}
\end{figure}

As such, this behaviour for the perturbed case is expected for a superfluid, for which substantial dissipation should only set in when the fluid velocity relative to the stirrer surpasses a certain critical treshold velocity. For our protocol, which starts at zero velocity (see also the second panel of \autoref{fig:data1}), this can only happen after a certain driving time. To examine this in more detail, we plot in \autoref{fig:Energies} the same internal energy, together with the total energy and COM energy. We see that initially, nearly all the energy that is injected into the system goes into the COM motion, $\Delta e \approx \Delta e_{COM}$, implying a vanishing dissipation rate $\dot e_{int}\approx 0$. After $t= t_{onset}\approx 50$~ms the dissipation sets in. The system then soon enters a quasi steady-state regime, in which - up to quasi-periodic modulations - the energy injection is counterbalanced by the dissipation, leading to a quasi-constant dissipation rate $\dot e \approx \dot e_{int}$ and a quasi-constant COM energy $e_{COM}$. 

The red triangle and red cross show two estimates of $t_{onset}$ based on the local Landau criterion. As shown in \cite{Albert2008}, in the case of a positive longitudinal stirrer $\delta V(y)>0$ that crosses the entire harmonic trap, application of the local Landau criterion in an effective 1D perturbative approximation for cigar-shaped traps in the 3D Thomas-Fermi regime, leads to a critical COM velocity, \be v_c=\left(1- \frac{\delta V_0}{\mu_0}\right)^{5/2}c_0\,, \label{vcrit}\ee where $c_0=\sqrt{\mu_0/(2m)}$ is the longitudinal sound-velocity at the center of the unperturbed cloud. The red triangle shows the corresponding analytical estimate for $t_{onset}$, as the first time the undamped driven COM velocity, $v(t)$ in Eq. (\ref{eomunpert}), surpasses this critical velocity. Our other estimate (red cross) relies on our 3D mean-field simulations and is obtained by applying the local Landau criterion on the simulated mean-field, as described in \autoref{sec:mean-field}. As can be seen in \autoref{fig:Energies}, the two estimates differ slightly. Nevertheless, we find that the local Landau criterion, although it assumes a constant fluid velocity - which is manifestly not the case in our setup - remains a good qualitative predictor for the onset of dissipation.

\begin{figure}[t]
\centerline{\includegraphics[width=1.05\columnwidth]{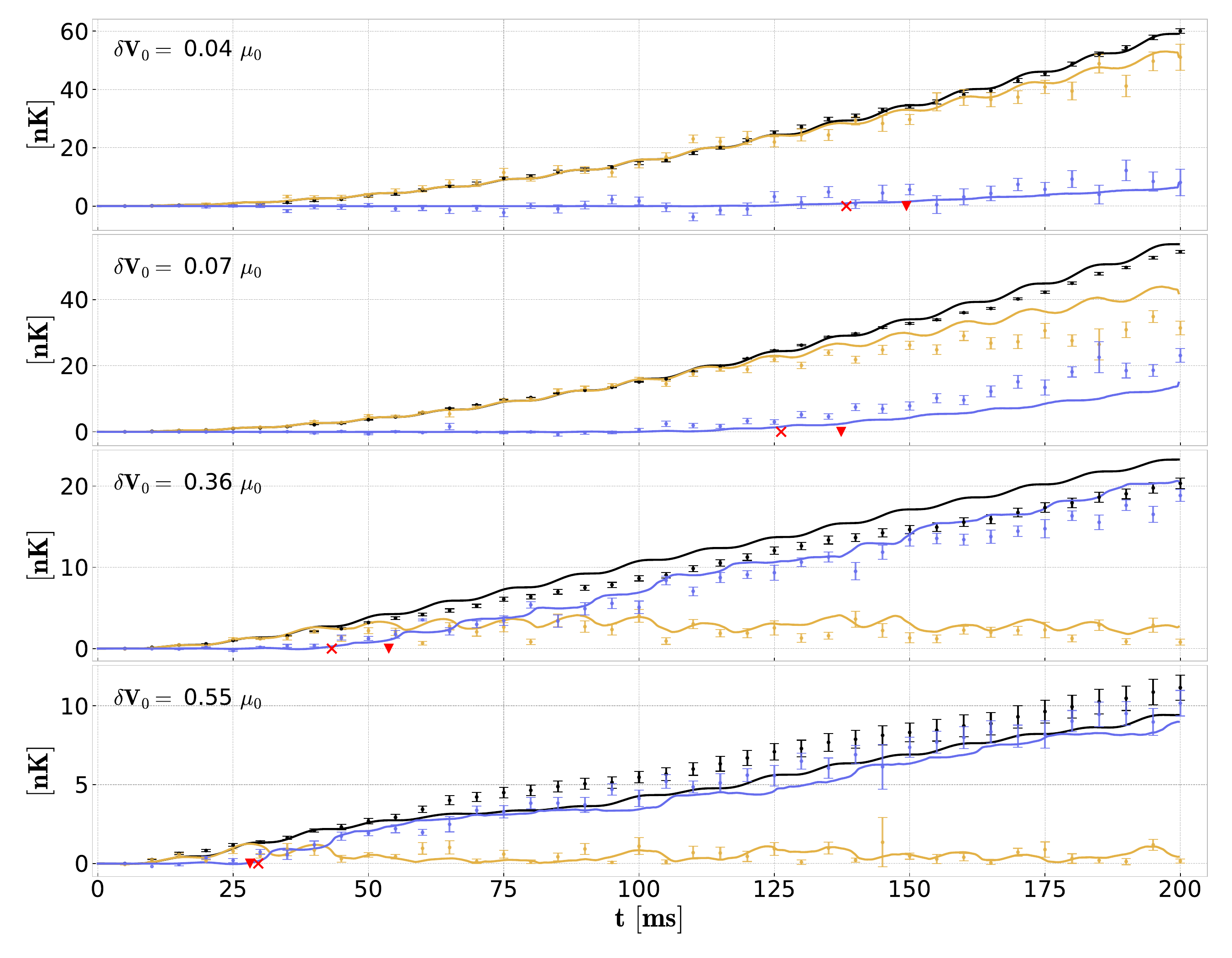}}
    \caption{\justifying \textbf{Energy curves for different stirrer strengths.} As in \autoref{fig:Energies}, but now for different stirrer strenghts $\delta V_0/\mu_0=0.04,0.07,0.36,0.55$. Notice the different vertical scale for each stirrer strength $\delta V_0$. }
    \label{fig:data2}
\end{figure}

This is also the case for other stirrer strengths $\delta V_0$, as shown in  \autoref{fig:data2}. For larger values of the potential, the dissipation sets in at earlier times, which corresponds to a smaller critical COM velocity, in line with the prediction (\ref{vcrit}), and in line with other earlier work \cite{Dries2010, Kwon2015,Kiehn2022, Pavloff2002, Watanabe2009, Huynh2022, Kwak2023}. Notice also that for $\delta V_0/\mu_0=0.34, 0.55$, we find a qualitatively similar quasi steady-state regime of dissipation as we see for the $\delta V_0/\mu_0=0.27$ case, but with different (approximately) constant rates $\dot e\approx \dot{e}_{int}$. For the two smaller stirrer strengths we do not observe such steady-state regime within the experimental drive times. Also our mean-field results, simulated up to longer drive times $t_{drive}=400$ ms (not shown) do not exhibit a quasi steady-state regime for those smaller stirrer strengths. Remark that for a positive stirrer strength, as a function of the stirrer velocity $v$, the frictional force (drag) $f'(v)$ and corresponding dissipation rate $\dot e_{int}(v)=f'(v).v$ is expected to peak around the critical velocity, with an increasing peak height for growing stirrer strength, see e.g the analytical results of \cite{Pavloff2002, Astrakharchik2004} and numerical results of \cite{Feng2019,Kiehn2022}. Our results indicate that for the three larger stirrer strengths this peak height is sufficient to balance out the driving force $f(t)$ in the time-averaged sense (over one driving period). In contrast, for the two smaller stirrer strengths, the peak height seems insufficient, implying the absence of a steady-state regime.

\begin{figure}
\centerline{\includegraphics[width=1.0\columnwidth]{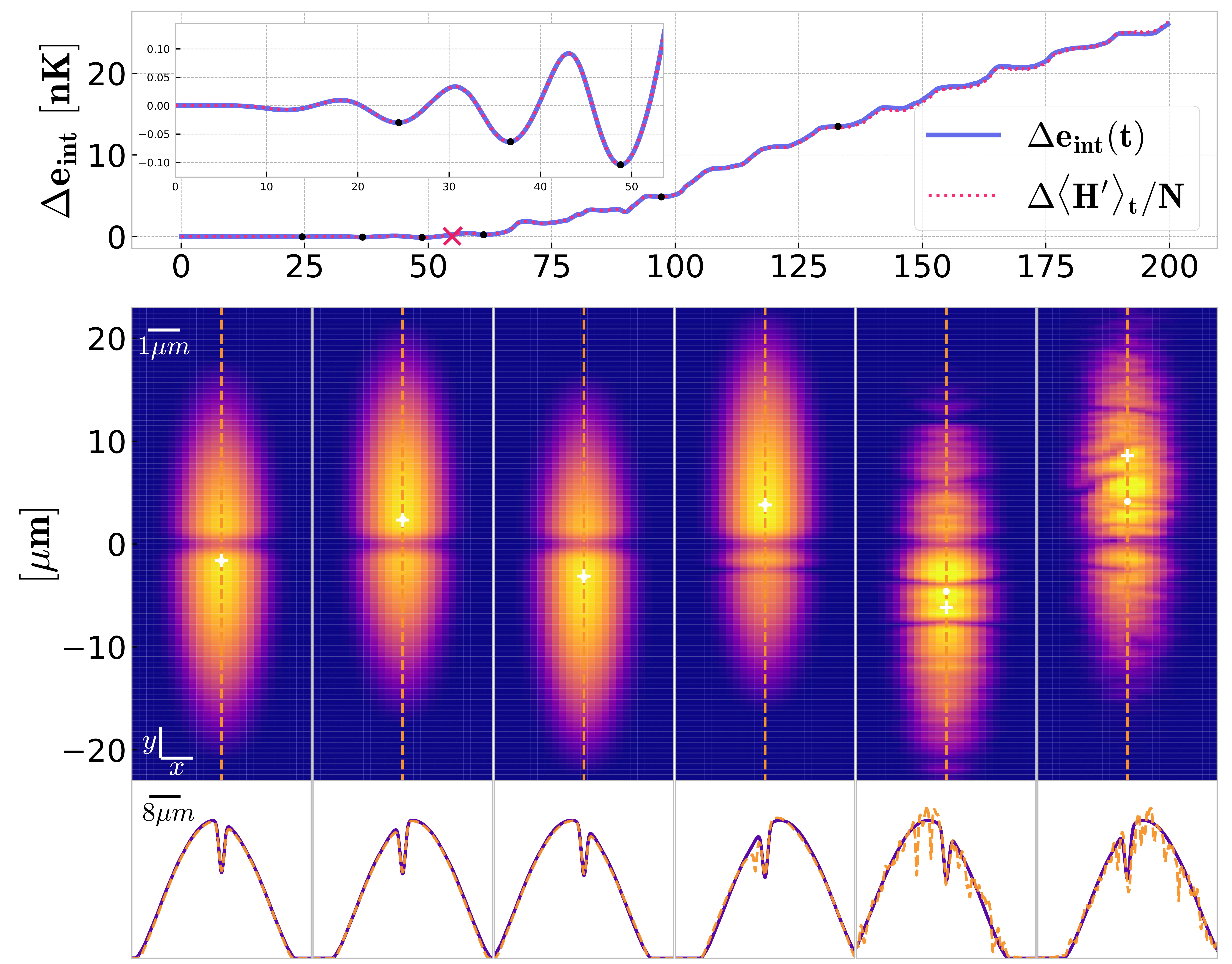}}
    \caption{
    \justifying
    \textbf{Mean field simulation.} 3D GPE simulation for the potential perturbation $\delta V_0 =0.27 \mu_0$. The upper panel displays the internal energy evolution, with the red cross indicating our mean-field estimate - via the local Landau criterion - of the onset time of dissipation. The middle panel shows column densities $n(x,y)$ at selected times, indicated by black dots in the upper panel. The white dots (crosses) indicate the COM position for the perturbed (unperturbed $\delta V_0=0$) cloud. The lower panel compares the cross-section along the central vertical cut of the simulated column density (orange) with the instantaneous groundstate column density (purple) in the COM rest frame.
    }
    \label{fig:INSITU}
\end{figure}

\section{Mean-field simulations} \label{sec:mean-field} We performed 3D simulations of the Gross-Pitaevskii equation (GPE) using a 4th order Runge-Kutta integrator in the interaction picture \cite{VortexDynamicsThesis}, taking $(31,701,31)$ lattice points in the $(x,y,z)$ directions. Overall the simulations show good qualitative agreement with the experimental data, possible causes for deviations are the uncertainties on the atom number and on the conversion of beam intensity to optical stirrer strength, which we both estimate to be of order $\pm 10\%$; and the omission of a (small) thermal component.

The upper panel of \autoref{fig:INSITU} shows the simulated internal energy evolution for $\delta V_0=0.27\mu_0$. As a sanity check on the perturbed HPT we have computed the internal energy in two different ways. The blue full line shows the computation from the COM position and velocity, employing the HPT energy conservation relation as in our experimental procedure. The dotted red line, overlapping perfectly with the first line, shows the direct computation from the COM rest frame Hamiltonian (\ref{Hcom}).

The red cross shows the estimate of the time $t_{onset}$ at which dissipation should set in, from applying the local Landau criterion on the simulated mean-field $\psi({\bf x},t)=\sqrt{n({\bf x},t)} e^{i\theta ({\bf x},t)}$, see also \autoref{fig:data2}. In particular we determine our mean-field estimate of $t_{onset}$ as the first time at which the local condensate velocity evaluated at the location of the potential (i.e. at the trap center), \be \frac{\hbar}{m} \partial_y \theta (\vec{x},t)\big\rvert_{\vec{x}=\vec{0}}\,,  \ee surpasses the local longitudinal sound-velocity $c_0(0,t)$ with (in the 3D Thomas-Fermi regime of \cite{Albert2008}):
\be c_0(y,t)=\left(\frac{\hbar \omega_{\perp}}{m}\right)^{1/2}\left(an(y,t)\right)^{1/4}\,,\label{1Dsound}\ee
where $\omega_\perp=\omega_{x/z}$, 
$n(y)=\int\int dx dz  n(x,y,z)$ the 1D density and $a$ the 3D scattering length.

Prior to the onset of dissipation $t_{onset}\approx $ 50~ms, our simulations show \emph{adiabatic flow} with the BEC tracking the instantaneous groundstate in the COM rest frame for the displaced potential $\delta V'$ in this frame, as illustrated by the first three snapshots in the middle and lower panel of the figure. For the internal energy evolution, this adiabatic behavior manifests itself by small oscillations - smaller than our experimental uncertainties -  that follow from the groundstate energy dependence on the stirrer position, see the inset in the upper panel. After $t_{onset}$ (last three snapshots) we clearly observe a qualitatively different behavior, with the production of soliton-like excitations, that in turn decay into smaller structures.

To get a better view on the excitations, we display in \autoref{fig:COMVIEW} this process in the COM rest frame ($y'=y-r_y(t)$, with $r_y(t)$ the COM position). For each of the depicted stirrer strenghts we see again the initial adiabatic flow - characterized by a sole moving underdensity at the stirrer position - and the subsequent production of excitations. We clearly find dark solitonic excitations, with large underdensities, small velocities and large jumps $\Delta \theta' \approx \pi$ of the condensate phase (in the COM rest frame) across the excitation; and what appears to be grey solitonic excitations with smaller underdensities, larger velocities $v_e\approx c_0$ and smaller phase jumps. In addition, the density plots also reveal small overdense excitations moving at the speed of sound, corresponding to sound-waves, see related findings in \cite{Hakim1997,Katsimiga2018, Feng2019}. Note that, especially for the larger stirrer strengths, our obtained values for $t_{onset}$ from the local Landau criterion - which in principle only considers dissipation into sound-waves - are actually also good estimates of the onset of soliton production, as was already observed in \cite{Feng2019} for a stirring protocol with constant velocity.

\begin{figure}
\centerline{\includegraphics[width=1.0\columnwidth]{Trajectories3.jpg}}
    \caption{
    \justifying
    \textbf{Production of excitations viewed in the COM rest frame.} For $\delta V_0/\mu_0=0.04,0.27,0.55$ the mean-field time-evolution of the 1D density in the COM rest frame. Red crosses: our mean-field estimates of $t_{onset}$ via the local Landau criterion. The initial paths of certain selected excitations are marked by dotted (dashed) lines for excitation velocities $v_e$ near the speed of sound $v_e=c_0 \pm 5\%$ (below the speed of sound $v_e<c_0 \pm 5 \%$), with $c_0$ the local longitudinal sound velocity (see Eq. (\ref{1Dsound})). White lines track underdensities (solitons), black lines track overdensities (phonons). The right-column shows the condensate phase profile $\theta'(0,y',0)$ (in units of $\pi$) along the blue lines indicated in the density-plots.}
    \label{fig:COMVIEW}
\end{figure}

\begin{figure}[t]
    \centerline{\includegraphics[width=\columnwidth]{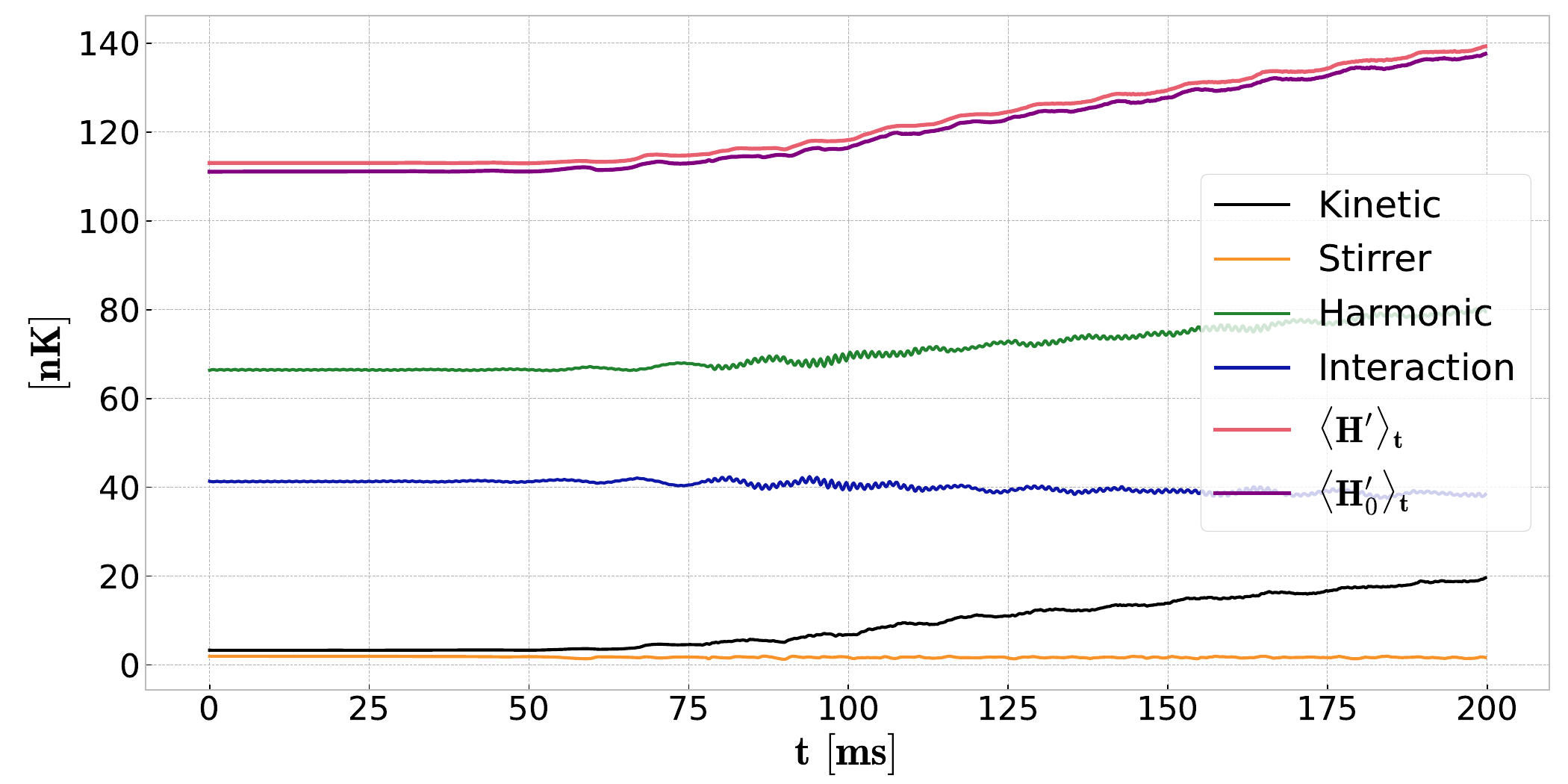}}
    \caption{
    \justifying
    \textbf{Internal energy contributions.}  For $\delta V_0=0.27 \mu_0$, mean-field results for the four different contributions to the total internal energy evolution (per particle): the kinetic, harmonic and interaction terms making up $H'_0$ and the stirrer term $H'_\delta$.}
    \label{fig:E_CONTRIBUTIONS}
\end{figure}

Finally, let us comment on one important caveat regarding our internal energy measurements. By construction the energy measured from the HPT energy conservation relation (\ref{econs}) includes the contribution from the stirring-term in the Hamiltonian:  $E_{int}=\braket{H'}=\braket{H'_0}+\braket{H'_\delta}$, while one is typically more interested in the energy of the system in the unperturbed trap, ${E_{int}}_0=\braket{H'_0}$. However, for a localized stirrer $\delta V$ we can expect the contribution of the stirrer-term to be subdominant to the other extensive terms in the Hamiltonian. This is indeed what we find in our simulations, as shown in \autoref{fig:E_CONTRIBUTIONS}.

\section{Conclusions} In this work, we have put forward a method for measuring the internal energy of linearly driven, uniformly interacting quantum many-body systems confined in perturbed harmonic traps. We demonstrated its experimental feasibility on a driven BEC, revealing energy dissipation curves characteristic of superfluid behavior. Looking ahead, it will be interesting to explore whether such internal energy measurements can yield spectroscopic insights into the nature of excitations (phonons, solitons, and vortices) generated during the dynamics. Other promising applications include the study of energy dissipation resulting from the two-component dynamics in finite temperature BECs \cite{ZNG1999} or energy transport in quantum turbulence, see e.g. \cite{Dogra2023}. In particular, in the latter context, it will be worthwhile to further investigate if and how our observed quasi-steady state regime gives rise to a non-thermal fixed point in the internal momentum distribution \cite{Sergey2024, Bagnato2025}.  

\newpage

\section*{Author contributions} All authors contributed equally to this work. CT and SVW set up the experiment, designed and executed the experimental protocols and the data-analysis, KVG performed the mean-field simulations and contributed to the theoretical results and the data-analysis, KVA conceptualized the project and supervised the experimental and numerical work. All authors discussed the different results and prepared the manuscript.   

\section*{Data availability} The experimental data as well as a Python Notebook containing code to process and analyze the data is available at \cite{data_zenodo}.

\acknowledgments{We thank Lucas Levrouw, Jacques Tempere, Peter Schlagheck and Ludwig Mathey for their comments, the Infleqtion team - in particular Evan Salim and Yewei Wu - for their technical support and Frank Verstraete for his material and moral support in establishing the cold atoms lab. This work is supported by FWO grants I004520N and  G0C5123N. CT also acknowledges support by the FWO PhD-grant 1176525N.}

\begin{widetext}
\section*{Appendix}
\appendix
\renewcommand{\sectionautorefname}{Appendix}

\section{The Hamiltonian in the COM frame}\label{sec:A1}
To derive the Hamiltonian $H'$ in the COM rest frame and the corresponding energy conservation relation \eqref{econs}, it is instructive to first consider an arbitrary frame transformation, ${\bf x}\rightarrow {\bf x}'={\bf x}-\tilde{\bf r}(t)$, for a general $\tilde{\bf r}(t)$ that does not necessarily coincide with the COM position. For the corresponding field transformation we write: \be \psi'({\bf x}',t)=e^{-\frac{i}{\hbar}(m \tilde {\bf v}. {\bf x}'+S(t))}\psi({\bf x},t)=e^{-\frac{i}{\hbar}(m \tilde {\bf v}. {\bf x}'+S(t))}\psi({\bf x}'+\tilde{\bf r}(t),t)\,,\ee where $\tilde{\bf v}=\dot{\tilde{\bf r}}$ is the instantaneous velocity of the new frame and we leave $S(t)$ free for now. Note that the first velocity-term in the phase factor follows from imposing the proper momentum-shift:
\be \hat p'_i=\frac{1}{N}\int\hspace{-0.13 cm}d^3x'{\psi'}^\dagger(-i\hbar\partial'_i)\psi'=\frac{1}{N}\int\hspace{-0.13 cm}d^3x\,{\psi}^\dagger(-i\hbar\partial_i-m\tilde{v}_i)\psi  =\hat p_i- m \tilde{v}_i \,.  \ee

For the time-evolution equation for $\psi'$ we then have:

\bea i\hbar\partial_t\psi'({\bf x}',t)&=&-\frac{\hbar^2}{2m}{\nabla'}^2\psi'({\bf x}',t)+[\dot S-\frac{1}{2}m|\tilde{ \bf v}|^2+\frac{1}{2}m\omega_i^2\tilde{r}_i^2-{\bf f}.\tilde{\bf r}]\,\psi'({\bf x}',t)+x'_i [m \dot {\tilde v}_i+m\omega_i^2 \tilde r_i-f_i]\,\psi'({\bf x}',t) \nn\\
&&+[\frac{1}{2}m\omega_i^2 {x'_i}^2+\delta V({\bf x}'+\tilde{\bf r})]\,\psi'({\bf x}',t)+\int\hspace{-0.13cm}d^3 y' V_I({\bf x}'-{\bf y}'){\psi'}^\dagger({\bf y}',t)\psi'({\bf y}',t)\psi'({\bf x}',t)\,,\label{eq:psi'}\eea
where we used the time-evolution of $\psi({\bf x},t)$ generated by the lab frame Hamiltonian $H$ (\ref{eq: H}). We now fix $S(t)$ such that the constant energy-shift (second term in right-hand side of Eq.\eqref{eq:psi'}) is canceled: 
\be S(t)=\int^t\hspace{-0.1cm} du\,\left( \frac{1}{2}m|\tilde{ \bf v}(u)|^2-\frac{1}{2}m\omega_i^2\tilde{r}_i^2(u)+{\bf f}(u).\tilde{\bf r}(u) \right)\,,\ee
which is the classical action for the frame motion $\tilde{\bf r}(t)$ in the driven harmonic trap. With this choice for $S(t)$ we identify the Hamiltonian $H'$ that generates the time-evolution Eq.\eqref{eq:psi'} in the moving frame:
\bea H'&=&\int\hspace{-0.13cm}d^3 x'\left( \frac{\hbar^2}{2m}{\bf \nabla'}{\psi'}^\dagger({\bf x}').{\bf \nabla'}\psi'({\bf x}') +[\frac{1}{2}m\omega_i^2 x'_i{}^{2}+\delta V'({\bf x}',t)-{\bf f}'(t).{\bf x}')]{\psi'}^\dagger({\bf x}')\psi'({\bf x}')\right)\nn\\&&+\int\hspace{-0.13cm}d^3 x'd^3y' \frac{1}{2}V_I({\bf x}'-{\bf y}'){\psi'}^\dagger({\bf x}'){\psi'}^\dagger({\bf y}'){\psi'}({\bf x}'){\psi'}({\bf y}')\,,\label{eq:Hprime} \eea
with: \be f'_i=f_i-m \dot {\tilde v}_i-m\omega_i^2 \tilde r_i\,,\quad\quad\textrm{and}\quad\quad \delta V'({\bf x}',t)=\delta V({\bf x}+\tilde{\bf r}(t))\,.\ee

Furthermore, by writing the integrand of $H'$ in terms of ${\bf x}$ and $\psi({\bf x})$ one can show the following relation between the lab frame Hamiltonian $H$ and the moving frame Hamiltonian $H'$:
\be H= H'+N(\hat{\bf p}.\tilde{\bf v}-m \hat{\bf r}.\dot{\tilde{\mathbf{v}}}-\frac{1}{2}m |\tilde {\bf v}|^2-{\bf f}.\tilde{\bf r}+m \dot{\tilde{\mathbf{v}}}.\tilde{\bf r}+\frac{1}{2}m \omega_i^2 \tilde{r}_i^2)\,, \ee
with $\hat {\bf r}$, $\hat{\bf p}$ the COM position and momentum operators defined in Eq.\eqref{eq:defrp}. Upon taking the expectation value we find the relation (with $e=\braket{H}_t/N,\,e'=\braket{H'}_t/N $, ${\bf r}=\braket{\hat{\bf r}}_t$, ${\bf v}=\braket{\hat{\bf p}}_t/m$):
\be e= \frac{1}{2}m \tilde {\bf v}.(2{\bf v}-\tilde{\bf v})+m \dot{\tilde{\mathbf{v}}}.(\tilde{\bf r}-{\bf r})+\frac{1}{2}m \omega_i^2 \tilde r_i^2-{\bf f}.\tilde{\bf r}+e'\,.
\label{e'cons}\ee

Finally, let us specify to the COM restframe: $\tilde {\bf r}={\bf r}$, $\tilde {\bf v}={\bf v}$. In this frame the transformed drive and potential in the Hamiltonian $H'$ (\ref{eq:Hprime}) read:
\be f'_i=f_i-m \dot { v}_i-m\omega_i^2  r_i=\frac{1}{N}\int\hspace{-0.13cm} d^3x\,\langle \partial_i(\delta V)\psi^\dagger\psi\rangle_t\,,\quad\quad\textrm{and}\quad\quad \delta V'({\bf x}',t)=\delta V({\bf x}+{\bf r}(t))\,,\label{eq:Hprimebis}\ee
where the right-hand side for the first equation follows from the COM equations \eqref{eomr} and \eqref{eomp}. And specifying the relation \eqref{e'cons} to the COM rest frame yields our energy conservation relation \eqref{econs}: 
\be e= \frac{1}{2}m |{\bf v}|^2+\frac{1}{2}m \omega_i^2  r_i^2-{\bf f}.{\bf r}+e_{int}=e_{COM}+e_{int}\,, \ee
where we now denote $e'=e_{int}$ for the energy in the COM rest frame and $e_{COM}$ indeed corresponds to the classical energy associated to the COM motion in the driven harmonic trap.

\section{Experimental set-up}
\label{section: experimental set-up}
The heart of our system consists of a two-chamber ultrahigh vacuum system (Infleqtion RuBECi). 
A Rubidium vapor is released from a dispenser in the lower chamber where a two-dimensional magneto-optical trap ($2D^+$-MOT) then feeds a three-dimensional MOT in the upper chamber \cite{FarkasRuBECi2010, Salim1Hz2014}. 
The magnetic quadrupole field for the $3D$-MOT is created by two coils located along the $x_{lab}$-axis on either side of the upper chamber -- the MOT-X-coils -- run in anti-Helmholtz formation. Two pairs of coils, along the $y_{lab}$- and $z_{lab}$-axes, are set in the Helmholtz configuration to generate homogeneous bias fields.
The upper chamber is closed off by an atom chip of thickness 420~$\mu$m, in the center of which a 3~mm diameter polished glass window is incorporated \cite{SalimChipWindow2013, AtomChipReview2002}. This window allows for optical access to project extra optical potentials as well as in-situ imaging.
See \autoref{fig: Atom Chip} for a schematic drawing.
On the vacuum side of the atom chip, 100~$\mu$m-wide traces -- the guide wires -- lie at a distance $x_{lab}=\pm$ 250~$\mu$m from the center of the window. We use one or both of these inner wires, together with an orthogonal bias field, to create $(x,z)$-radial confinement.
The T-shaped traces on the atom chip were not used for the experiment described in this paper.
On the ambient side, 400~$\mu$m wide wires at distance $y_{lab}=\pm$ 1.25~mm from the center of the window -- the H wires -- are used to close off the trap in the longitudinal ($y$-)direction.
\begin{figure}[h!]
    \centering
      \includegraphics[width=0.5\textwidth]{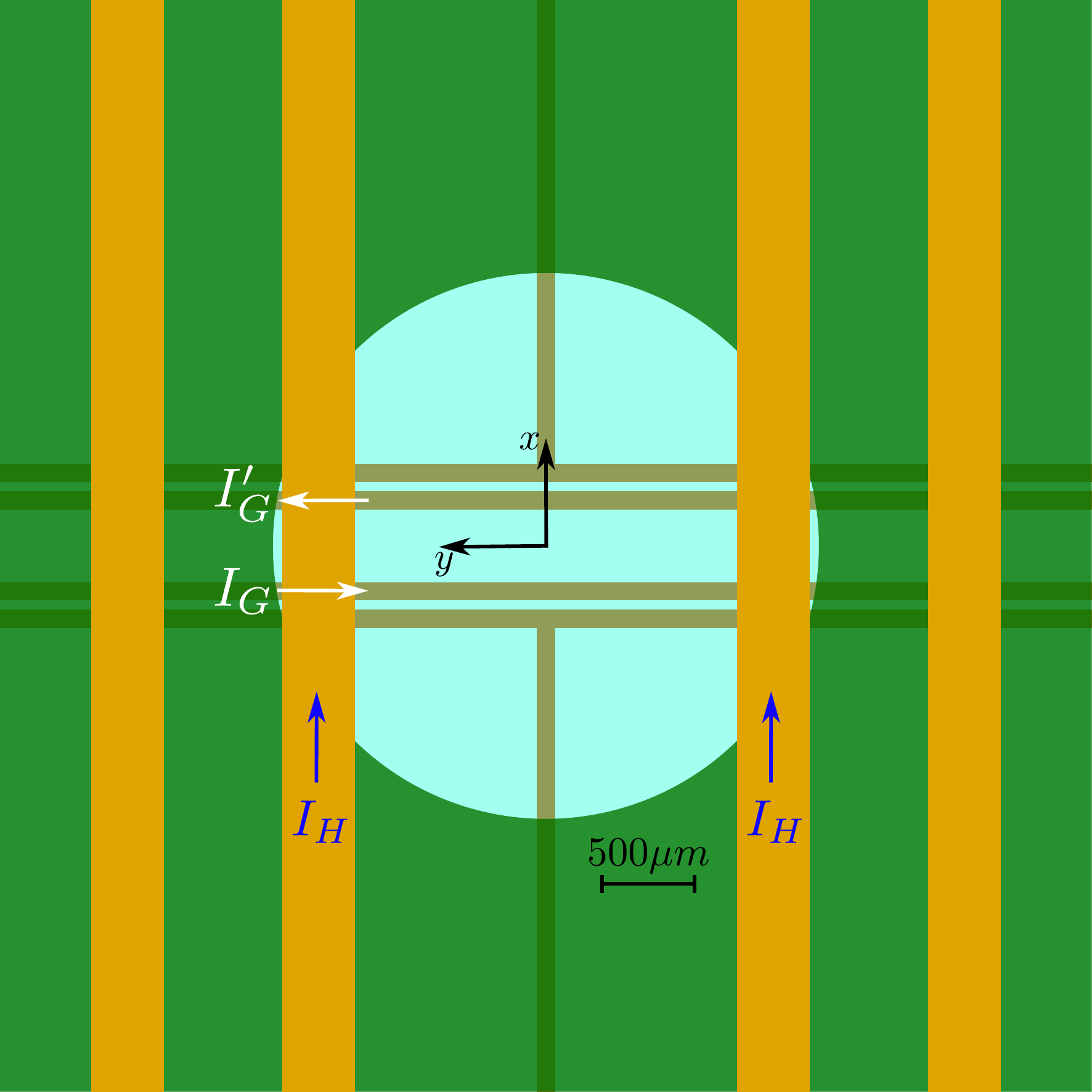}
    \caption{
    \justifying
    \textbf{Schematic view of the atom chip.} The lab coordinate system is centered on the window. The direction of the currents are indicated by the colored arrows.}
    \label{fig: Atom Chip}
\end{figure}
\\

A typical experimental run goes as follows, similar to \cite{Salim1Hz2014}. After loading $^{87}Rb$ atoms in a $3D$-MOT in the upper chamber, the cloud is spatially compressed, after which a short stage of polarization gradient cooling brings the atom cloud to sub-Doppler temperatures.
The atoms are then optically pumped into the $\ket{F = 2, m_F=2}$ ground state. A magnetic quadrupole trap is switched on to catch the atoms.
The cloud is transported toward the atom chip by linearly ramping the currents from the MOT-X-coils to transport-X-coils, mounted $\sim$ 1.5~cm higher towards the atom chip. Meanwhile, the bias fields, the counter-propagating currents in the guide wires ($I_G$ and $I_G'$), and the co-propagating currents in the H-wires ($I_H$) are turned on and adjusted to fully load the atoms in the harmonic chip trap about 350~$\mu$m below the window. The atom cloud is then transported to $\sim$ 90~$\mu$m directly below the lower guide wire ($G$), yielding a tight harmonic trap of (calculated) trap frequencies $(\omega_x, \omega_y, \omega_z) = 2\pi\times(1.8, 0.044, 1.8)$~kHz. Forced radio-frequency (RF) evaporation brings the atom cloud to just above degeneracy. 
The trap is again moved to the center of the chip-window, now about 145~$\mu$m below the chip surface, with (measured) trapping frequencies of $(\omega_x, \omega_y, \omega_z) =2\pi\times (440(2), 41.6(2), 440(2))$~Hz.
A final RF evaporation cools the cloud below 150~nK to a nearly pure BEC of $\sim55\,000$ atoms.
The atom cloud is shaped like an elongated cigar whose longitudinal $y$-axis is slightly misaligned by an angle $\theta\sim3\degree$ with respect to the $y_{lab}$-axis of the lab-frame, as shown in \autoref{fig: BiotSavart Imaging Trap}: 
$e_y=\cos{\theta} e_{y_{lab}} + \sin{\theta} e_{z_{lab}}$.
This is also visible in the time-of-flight images where the (shape-inverted) cloud is slightly rotated with respect to the camera alignment, see \autoref{fig: tofimages} in the main text.
The camera is set up according to the lab-frame and thus the measured horizontal displacements used in Eq.\eqref{eq: y(tof)} are not exactly along the atoms' principal axis $e_y$. However, since the rotation is small enough -- $\cos{\theta} \approx 1$ -- the longitudinal axes effectively match within the resolution of the imaging system.
\begin{figure}[ht!]
    \centering
    \includegraphics[width=\textwidth]{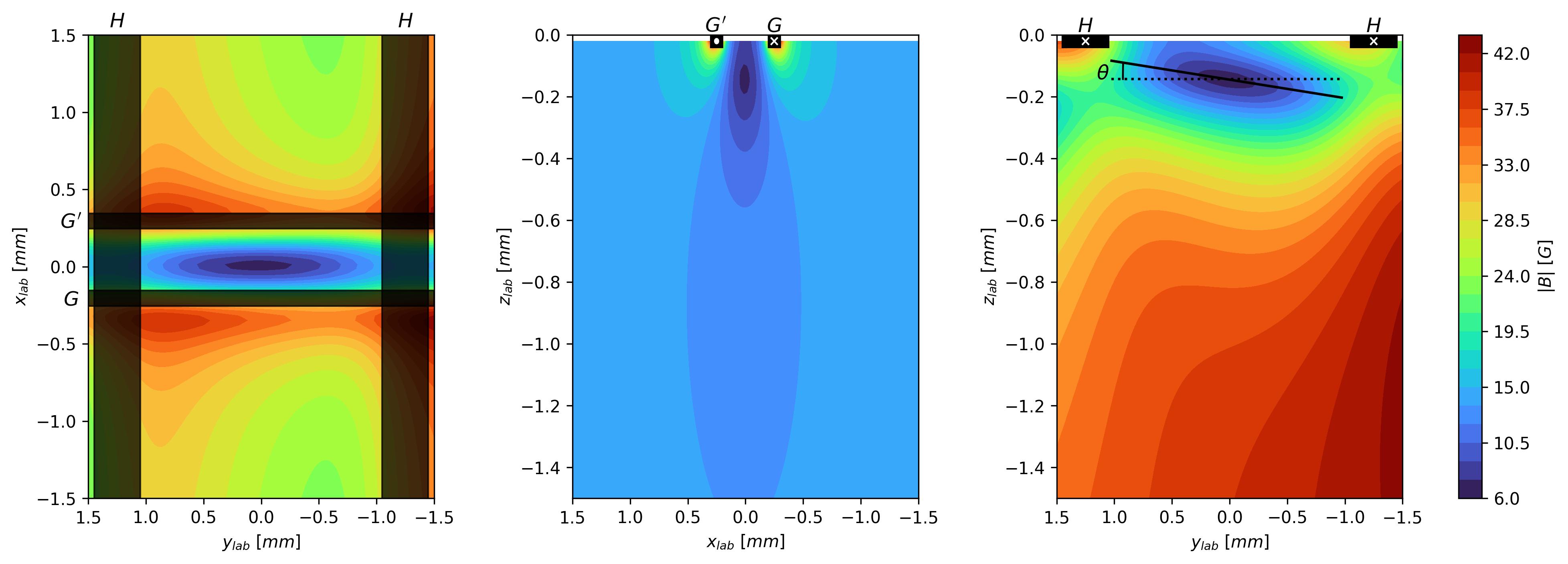}
    \caption{
    \justifying
    \textbf{The final trap geometry in which further experiments are performed.} The magnetic fields are calculated using the Biot-Savart law for infinitesimally thin wires of finite length and width.}
    \label{fig: BiotSavart Imaging Trap}
\end{figure}
\\

After loading the MOT in the upper chamber, it takes just less than 2.3 seconds to transport and cool the atoms to degeneracy in the center of the atom chip. The experiment, with variable duration, starts hereafter as explained in the main text and in appendix C. 

The projection light used for the stirrer is generated by free-running a tapered amplifier (Toptica BoosTApro) which seeds a second tapered amplifier (TA) (Toptica BoosTApro).
The result is a broadband laser beam peaked at 760 nm which is coupled into the optics mounted on a XY-translation platform above the upper vacuum chamber. Two perpendicular AODs (Gooch \& Housego AODF 4055-4) are used to shape the laser light into the desired potential by appropriate driving signals, which are converted by an Arbitrary Waveform Generator (National Instruments two-channel PXIe-5433). After passing the AODs, the output light field is relayed through telescope optics, passing a coaxial illuminator which directs the light to the microscope objective (MY20X-824 - 20X Mitutoyo Plan Apochromat Objective, $0.40$ NA, 20.0~mm $WD$), focusing the light onto the $XY$-atom plane. The coaxial illuminator also allows for in-situ absorption imaging in the same $z$-direction. For this, probe light is directed through a pinhole in the bottom of the lower vacuum chamber through the atom plane, passing through the coaxial illuminator, the microscope objective and is focused onto a sCMOS sensor (Oxford Instruments Andor Zyla 4.2 using the Infinity Photo-Optical KC VideoMax lens system). 
For this work, the in-situ imaging setup was used to calibrate the focus of the objective on the atom plane in between several experimental runs, to measure the intensity profiles, and to calibrate the conversion from beam intensity to potential (see below).

The default mode of a single tone driving in both AODs, yields a focused elliptical beam (spot) with waists:  \be w_x(z)= w_0\sqrt{1 + \left(\frac{z}{2.2 \mu m}\right)^2},\quad w_y(z)=w_0 \sqrt{1 + \left(\frac{z}{3.7 \mu m}\right)^2} \,,\ee with $w_0=$~1.2~$\mu$m,  where the values were measured from the reflection on the chip window. For the experiments in this paper, the perturbation is created by applying a sum of sines and a single-frequency sine wave for, respectively, the $x$ and $y$ directions. This results in an intensity profile (sum of spots): 
\bea 
I(x,y,z) &=& I_0\frac{w_0}{w_y(z)}e^{-\frac{2 y^2}{w_y(z)^2}}\sum_i \frac{w_0}{w_x(z)} e^{-\frac{2(x-x_i)^2}{w_x(z)^2}} \hspace{0.7cm} x_i \in \{-17, -16, \hdots, 17\}\mu m \nn\\
&\approx& \sqrt{\frac{\pi}{2}}1.2 I_0\frac{w_0}{w_y(z)}e^{-\frac{2 y^2}{w_y(z)^2}} \,,\label{eq:lijn}
\eea
We have also verified the latter intensity profile by direct measurement. The optical potential is directly proportional to this intensity, $\delta V(x,y,z)=C_d I(x,y,z)$, see Eq. (\ref{eq:pot}).  

The calibration of the conversion factor $C_d$ was done on in-situ images of the stationary BEC, resulting from evaporation in a trap with the projected line (\ref{eq:lijn}). We determined the smallest value of the amplitude $A$ of the single-tone AOD driving signal for which the atom density at the trap center $(0,0,0)$ is reduced to zero. This can be converted to a potential $\delta V_0\approx \mu_0$ (Thomas-Fermi regime). As the AODs operate in the linear regime, with the beam intensity quadratically dependent on the signal amplitude, $I\propto A^2$, this effectively gives us the intensity (AOD signal amplitude) to potential conversion factor.

To gather the time-of-flight data relevant for this experiment, standard time-of-flight absorption imaging is used. All magnetic fields are switched off and the atom cloud falls and expands during 24~ms. The probe light is flashed onto the atoms for 35~$\mu$s and into a CMOS camera-sensor (Basler acA2040-90umNIR using the Infinity Photo-Optical KX InifiniMax lens system).
A background shot of the probe light without atoms is taken 50~ms later.

\begin{figure}[t!]
    \centerline{\includegraphics[width=0.75\textwidth]{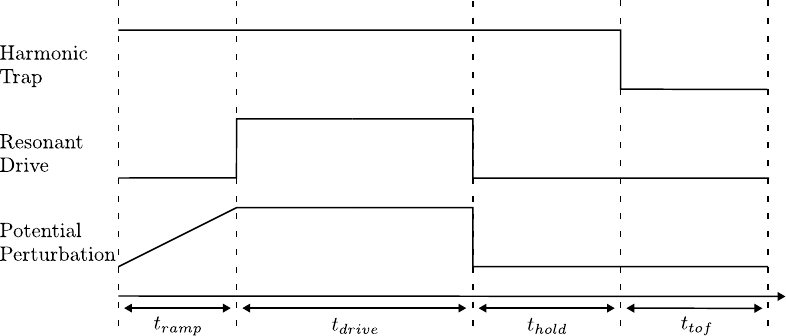}}
    \caption{
    \justifying
    \textbf{Experimental sequence}: the experimental sequence starts by ramping the perturbation linearly from zero to its final intensity over a duration $t_{ramp}=$20~ms: $\delta V_0(t)=\delta V_0 \times t/t_{ramp}$. In the second stage the resonant drive is switched on for $t_{drive}$ which varies between 5-200~ms. After this driving period, both the resonant drive and perturbation are switched off and the condensate is allowed to oscillate freely in the trap for a hold time $t_{hold}$, ranging from 0 to 25~ms. Finally, the trap is switched off, and after 24~ms of time-of-flight, the condensate is imaged in the $YZ$-plane via resonant absorption.
    }
    \label{fig: experimental sequence}
\end{figure}

\begin{figure}
     \centering
     \begin{subfigure}[b]{0.49\linewidth}
         \centering
         \includegraphics[width=\textwidth]{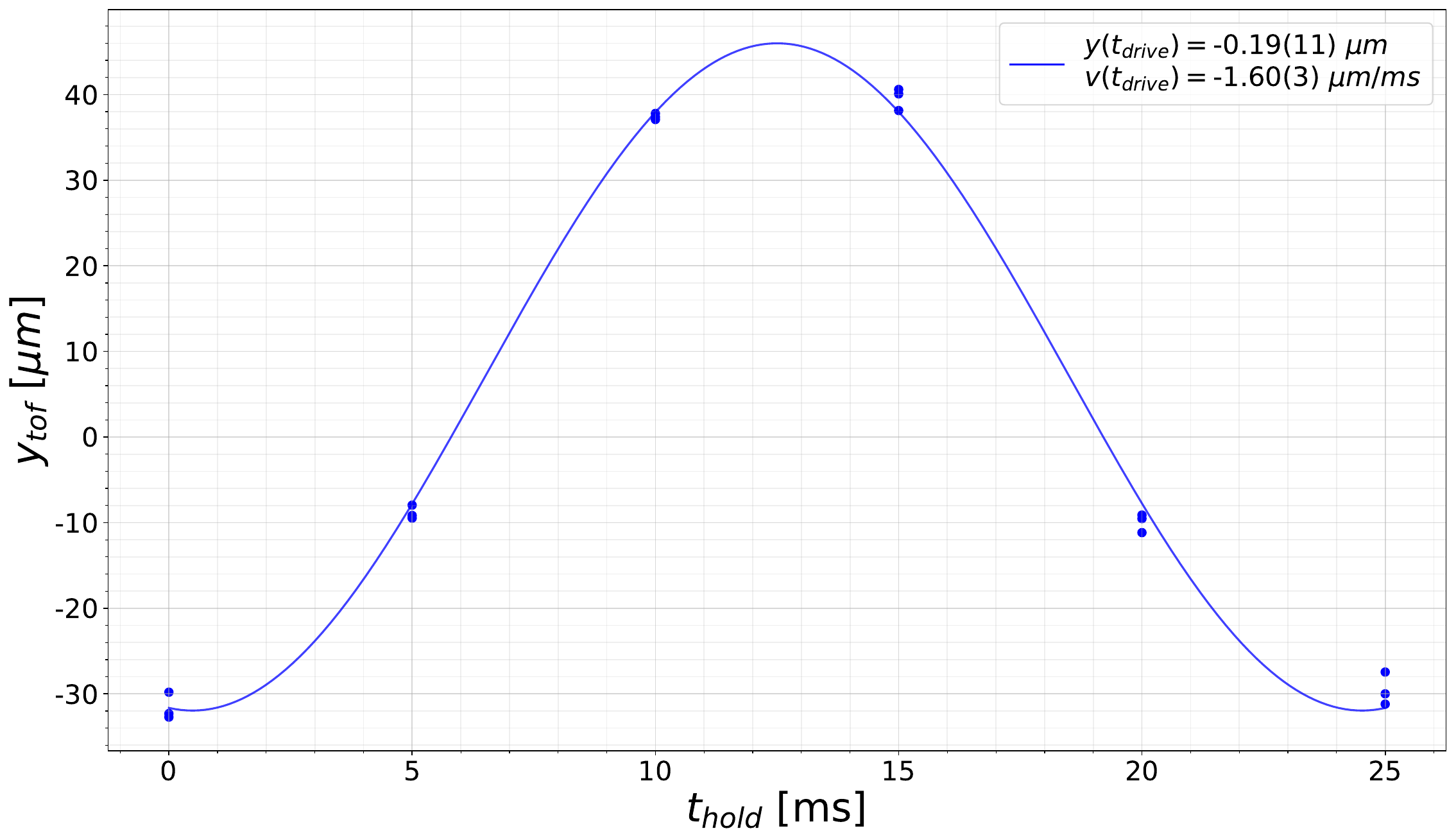}
         \caption{}
         \label{}
     \end{subfigure}
     \hfill
     \begin{subfigure}[b]{0.49\linewidth}
         \centering
         \includegraphics[width=\textwidth]{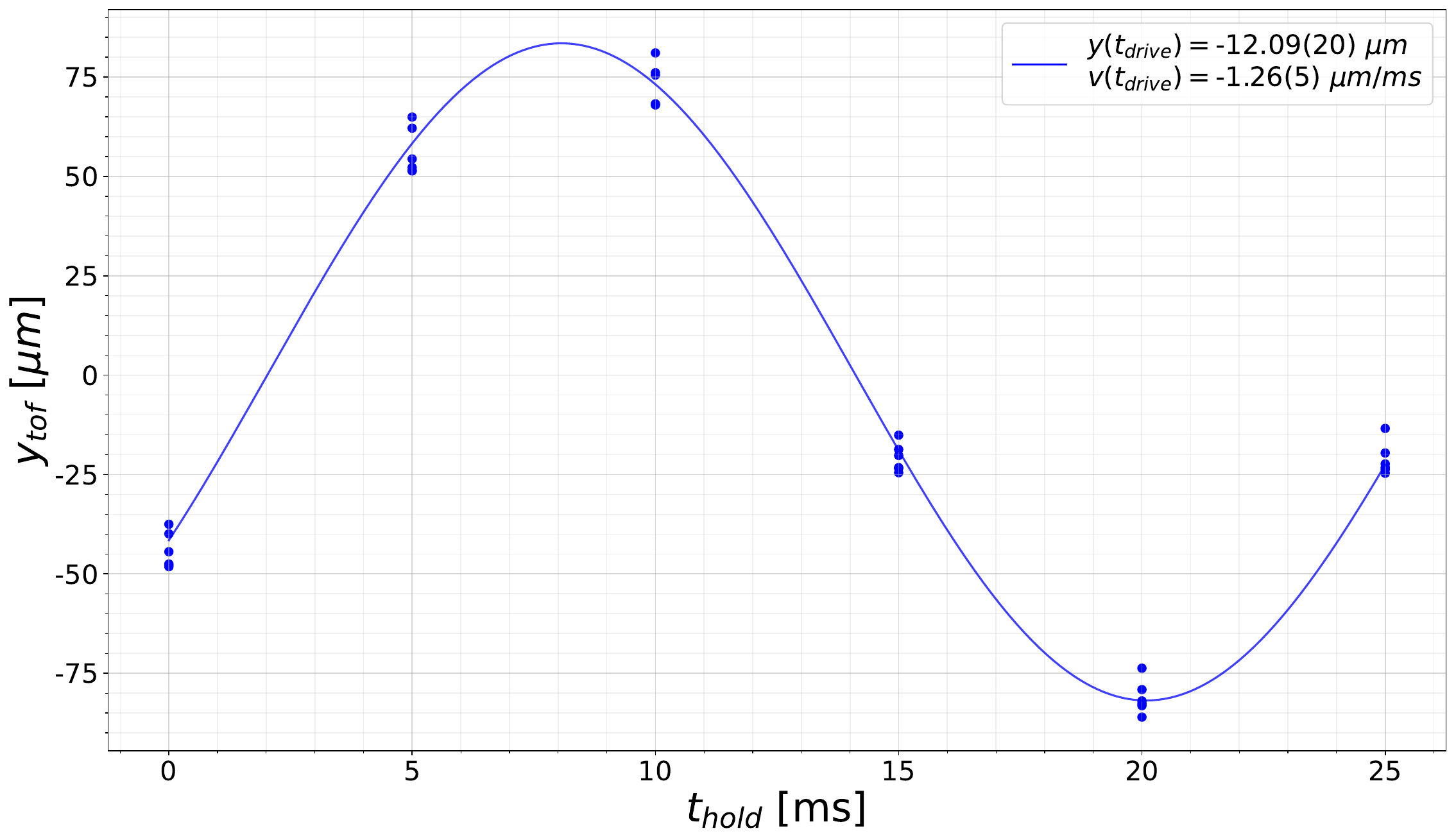}
         \caption{}
         \label{}
     \end{subfigure}
     \caption{\justifying
    \textbf{Obtaining $y(t_{drive})$ and $v(t_{drive})$ from the unperturbed HPT.} Fits of the time-of-flight positions to Eq.\eqref{eq: y(tof)}, to obtain the in-situ position and velocity after a specific value of $t_{drive}.$ The parameters are ($t_{drive}=$~90~ms, $\delta V_0=0.07\mu_0$) and ($t_{drive}=$~190~ms, $\delta V_0=0$) for (a) and (b) respectively. }
    \label{fig: example thold}
\end{figure}

\section{Position and velocity measurements}

\label{sec: Position and velocity measurements}

\autoref{fig: experimental sequence} shows our experimental protocol that we use to measure the in-situ COM position $y(t)$ and velocity $v(t)$ after resonantly driving the cloud in the perturbed trap for a certain drive time $t=t_{drive}$. By switching off the resonant drive and the perturbation after the driving sequence, according to the unperturbed HPT (Eqs. {\eqref{eomr},\eqref{eomp}}), this in-situ COM position and velocity will act as the initial conditions for an undamped harmonic motion during the hold stage:
\begin{equation}
\label{eq: eom harm osc}
\begin{aligned}
    y_{IS}(t_{hold}) &= y(t_{drive}) \cos(\omega_y t_{hold})+\frac{v(t_{drive})}{\omega_y}\sin(\omega_y  t_{hold})\,, \\
    v_{IS}(t_{hold}) &= -\omega_y  y(t_{drive}) \sin(\omega_y t_{hold})+v(t_{drive})\cos(\omega_y t_{hold})\,.
\end{aligned}
\end{equation}

Specifically, we have taken measurements for 6 equally spaced hold times, $t_{hold}=$0~ms, 5~ms,\ldots, 25~ms. Note that the maximal hold time $t_{hold}=$25~ms is chosen to be (slightly) larger than the period of the harmonic motion in the $y$-direction: $T={2\pi}/{\omega_y}=$~24.04~ms. After the hold stage we determine the horizontal COM position in time-of-flight after $t_{TOF}=$~24~ms: \be y_{TOF}=\frac{\sum_{ij}n_{(i,j)}y_i}{\sum_{ij}n_{(i,j)}}\,,\ee with $n_{(i,j)}$ the measured column density and $y_j$ the horizontal coordinate for a certain pixel $(i,j)$. This time-of-flight COM position is related to the in-situ position by: \begin{equation}
    y_{TOF}(t_{hold}) = y_{IS}({t_{hold}}) + v_{IS}({t_{hold}}) \cdot t_{TOF} \label{eq: y(tof)}\,.
\end{equation}

Note that the second term here shows the magnification of the COM motion in time-of-flight, which effectively enhances the resolution on the in-situ COM position and velocity. For all our data-points $(y(t_{drive}),v(t_{drive}))$ we have taken 3 TOF measurements for each $t_{hold}$, except for the unperturbed data with  $t_{drive} \in [100, 200]$, where in order to reduce the measurement errors we have taken 6 TOF measurements per hold time. For a certain value of $\delta V_0$ we have taken our measurements for all possible values of $(t_{drive},t_{hold})$ and all repetitions in randomized order. 
For a given value of $t_{drive}$ (and $\delta V_0$), the 18 (36) measured time-of-flight positions were then fitted to Eq.\eqref{eq: y(tof)} and Eq.\eqref{eq: eom harm osc}   with ordinary least squares, to produce a data-point $(y(t_{drive}),v(t_{drive}))$, as illustrated in \autoref{fig: example thold}.  
The covariance matrix $C_{ij}$ of this fit was used to estimate the variances on the position $\sigma_{y(t_{drive})}=C_{11}$ and velocity $\sigma_{v(t_{drive})}=C_{22}$, further used in the error-propagation for our determination of $e_{int}(t)$. Notice that in particular our determination of $\Delta e(t_j)=-\int^{t_j}_0\! du\, \dot f(u) y(u)$ depends on the position measurements at previous drive times $t_{drive}\leq t_j$. To evaluate this integral we have approximated it by the composite Simpson's 1/3 rule with the considered drive times $t_i$ as evaluation points. In the following section we motivate the viability of this approximation.

\section{Integration error estimation}\label{Sec: Integration error estimation}

\begin{figure}[t!]
    \centerline{\includegraphics[width=1\columnwidth]{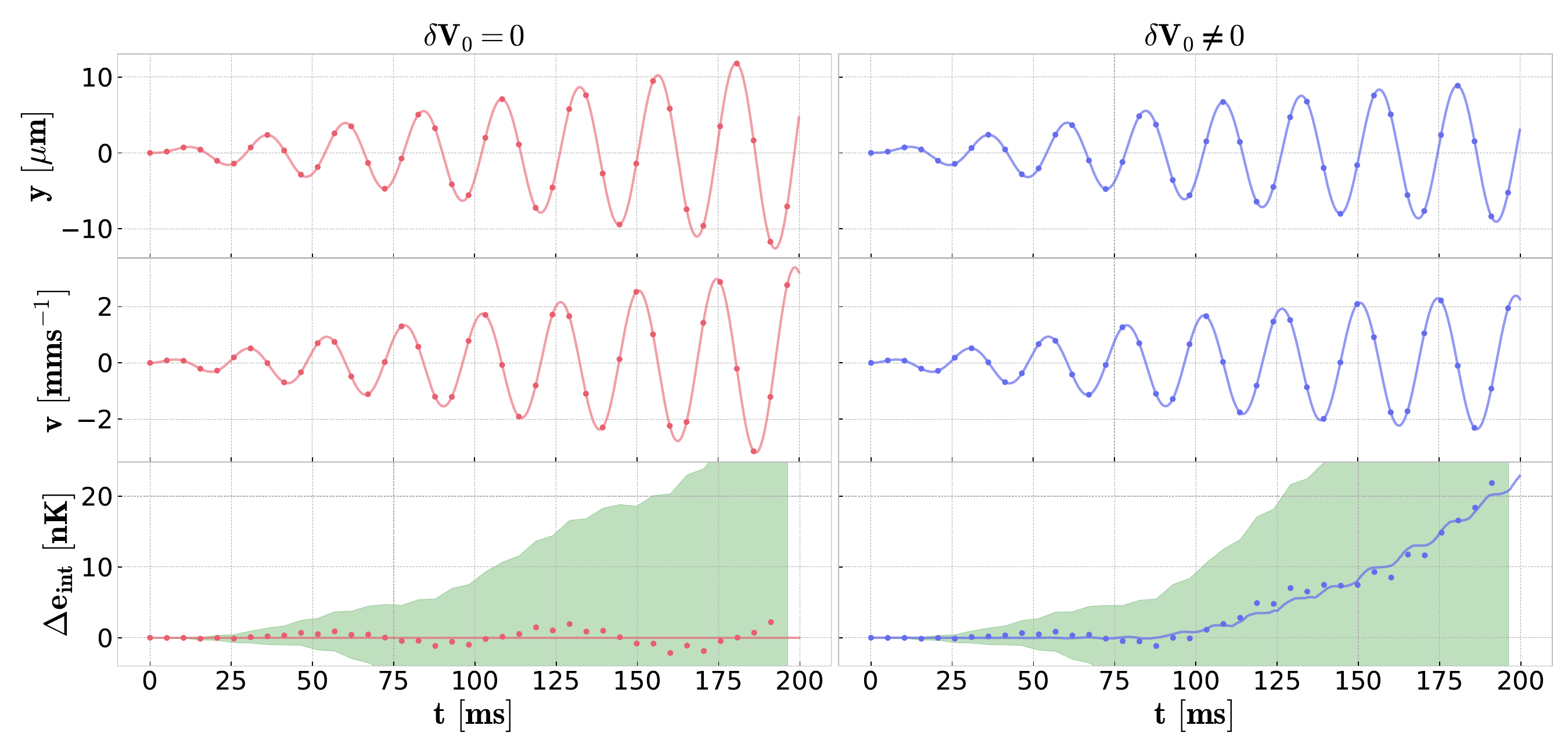}}
    \caption{\justifying\textbf{Integration error}: Internal energy calculation from sampled data-points $y_j$, $v_j$ on mean-field simulations for both $\delta V=0$ and $\delta V\neq0$ with sampling interval $\Delta t=5ms$. The deviation from the exact $\Delta e_{int}(t)$-curve is entirely due to the Simpson approximation. The green shaded areas show the upper bounds on the integration error (\ref{simpbound}).}
    \label{fig: ERROR}
\end{figure}
\label{sec: Simpson}
The internal energy given by Eq.\eqref{intenergyexp}: 
\be 
\Delta e_{int}(t) = \Delta e(t)-\Delta e_{COM}(t)= -\int_0^t du\dot{f}(u)y(u) - \left(\frac{1}{2}m v^2(t) - \frac{1}{2}m\omega^2 y^2(t) + f(t)y(t)\right)\,,
\ee 
is determined experimentally by approximating the integral expression for the total energy in the first term:
\be 
\Delta e_{int}(t_j)\approx \Delta \tilde e_{int,j}= -\frac{\Delta t}{3}\sum_{i=1}^{j}w_i \dot{f}_i y_i - \frac{1}{2}mv_{j}^2 - \frac{1}{2}m\omega^2y_j^2 + f_j y_j,\hspace{0.75cm} t_j = j  \Delta t,\hspace{0.75cm}j\geq 1\,,
 \label{eintapprox}\ee
with $\Delta t=$~5~ms, $w_i$ the weights of the composite Simpson's 1/3 rule and $y_j, v_j$ respectively denoting the position and velocity measurements. An upper bound for the integration error due to Simpson's approximation is given by:
\begin{equation}
|\Delta e_{int}(t_j)- \Delta \tilde e_{int,j}| \leq \frac{\Delta t^4}{180}t_j\max_{\xi \in [\Delta t,t_j]} |\frac{d^4 (\dot{f}y)}{dt^4}(\xi)|\,,
\label{simpbound}
\end{equation}
but as is well known, in practice this bound drastically overestimates the actual error and is therefore not useful for error estimation. To estimate the error in our case, we have compared exact computations of $\Delta e_{int}(t)$ for our numerical mean-field simulations, with the approximate result \eqref{eintapprox} from sampling the numerically determined continous curves $(y(t)$, $v(t))$. Both for $\delta V_0=0$ and $\delta V_0 \neq 0$ we find deviations that are considerably smaller than the estimated upper bound (\ref{simpbound}) and also smaller than the other induced estimated errors from the experimental measurement uncertainties on $(y_i,v_i)$, which motivates the use of Eq.\eqref{eintapprox}, neglecting the integration error altogether. 
\newpage

\end{widetext}

\end{document}